\documentstyle[11pt,rafringe,twoside,makeidx,epsfig,graphicx]{article}
\makeindex

\def\edcomment#1{\iffalse\marginpar{\raggedright\sl#1\/}\else\relax\fi}
\reversemarginpar

\begin{document}
\title{Variability, Brightness Temperature, Superluminal Motion, Doppler Boosting, and Related Issues} 

\author{K. I. Kellermann} 
\affil{National Radio
 Astronomy Observatory, 520 Edgemont Road, Charlottesville, VA, 22903, USA}

\begin{abstract}
We review the observations of rapid flux density variations in compact
radio sources, and discuss the inverse Compton limit to the maximum
brightness temperature of incoherent synchrotron sources in comparison
with recent VLBA observations.  The apparent agreement of the
theoretical brightness temperature limit due to inverse Compton
cooling and the brightness temperatures observed by early VLBI
observations appears to have been fortuitous.  VLBA observations have
greatly improved the quality of the data, but many of the early issues
remain unresolved.
\end{abstract}

\section{Variability Issues}

Starting in the mid 1960s, it was becoming increasingly clear that
many flat spectrum radio sources exhibited very rapid flux density
variations (e.g., Sholomitskii 1965; Dent 1965; 
Pauliny-Toth \& Kellermann 1968).  This presented serious problems
for conventional synchrotron models as it appeared to require a huge energy
content and an excessive amount of inverse Compton cooling.

\subsection{The Saga of \Index{3C\,120}}

One of most variable sources observed in the early years was 
\Index{PKS\,0430+05} (Day et al.\ 1966) 
also known as \Index{NRAO\,182} (Pauliny-Toth et al.\ 1966),
which is now widely, but incorrectly, known as \Index{3C\,120}.  The
misunderstanding about the nomenclature of this source goes back to
the program of flux density and variability observations made with the
newly completed 140-ft radio telescope at 10, 6, and 2\,cm.  The
observing list for these programs was created by combining the source
lists which Ivan Pauliny-Toth and I had separately made from our
previous work at the 300\,ft (mostly norther sources) and Parkes
telescopes (mostly southern sources) respectively.  One source was
common to our two lists, but was called \Index{NRAO\,182} by Pauliny-Toth and
\Index{PKS\,0430+05} by myself.  Thinking that such a strong centimeter source
should have been included in the better known low frequency
catalogs, we searched in vain for a source at the same or nearby
position in the 3CR catalog (Bennett et al.\ 1962)

Our inspection of the original confusion limited 3C Catalog was, at
best, confusing.  The position (B1950) given in the NRAO VLA
Calibrator Manual is ${RA} = 04^{\rm h}30^{\rm m}31\fs 6$ and $Dec = +05\deg
14\arcmin59\farcs 6$.  The position of 
\Index{3C\,120} in the original 3C Catalog is
$RA = 04^{\rm h}29^{\rm m}32^{\rm s} \pm 4^{\rm s}$ and 
$Dec = +01\deg 55\arcmin \pm 8\arcmin$
which is more than 3 degrees away in declination.  However, due to
noise and confusion, the positions of some sources in the 3C Catalog
can be in error by one or more full lobe shifts of the interference
pattern.  At +5 degrees declination, $\Delta RA = 57^{\rm s}$ and $\Delta
Dec = 2\deg13\arcmin 59\arcsec$.  Applying a correction of one lobe shift in
each coordinate and allowing for an additional 2 sigma error in
declination $(16\arcmin)$ gives $RA = 04^{\rm h} 30^{\rm m} 29^{\rm s} \pm 4^{\rm s}$ and
$Dec = +05 \deg 12\arcmin \pm 8\arcmin$.  Although this is in reasonable
agreement with the modern VLA position, after allowing for a lobe
shift in both coordinates as well as a 2~sigma shift in declination, a
3C source can be found at almost any position in the sky.  So, there
is probably no connection between the original \Index{3C\,120} and the variable
superluminal source which has received so much attention over the
past three decades.  Nevertheless, just for our own bookkeeping,
somewhat frivolously we called our source``\Index{3C\,120}'' in order not to
favor either one of us with the NRAO or Parkes designation.

Subsequently, we discovered remarkably rapid flux density outbursts in
``\Index{3C\,120}'' with observed increases of as much as 1\,Jy per week at 2\,cm
(Pauliny-Toth \& Kellermann 1968).  We were able to fit the change in
peak flux density with frequency and time with a simple expanding
source model (Shklovskii 1965; van der Laan 1966; Pauliny-Toth \&
Kellermann 1966) with multiple outbursts each separated by about a
year.  We were excited by these observations, and rushed into
publication, having forgotten that we had essentially {\em invented}
the name \Index{3C\,120} for this source. Since then, numerous authors have
used this nomenclature in hundreds of subsequent publications (see
NED). Apparently no one ever went back to look for \Index{3C\,120} in the
original 3C or 3CR catalogs.  This is a good opportunity to set the
record straight. There is no \Index{3C\,120}!  

Curiously, no radio source
since, has shown the same simple behavior.  Rather most sources have
multiple outbursts that overlap in frequency and time so that there is
no unique decomposition into individual outbursts.  In spite of the
early success in interpreting \Index{3C\,120} and \Index{1934$-$63} 
(Shklovskii 1965), and
other sources, it isn't clear how to accommodate the expanding source
model within currently popular shock in jet models (e.g., Marscher \&
Gear 1985).  Unfortunately, there were no VLBI observations at that
time to study how the structure changed with time.  Were there really
multiple expanding components?  If so, were they coincident in space,
or did they propagate along a jet-like path?  What was the size and
expansion velocity of these transient outbursts?  These questions
partially motivated the development of the first NRAO-Cornell VLBI
system.  But, by the time the MkI VLBI system became operational in 1967,
the isolated outbursts in \Index{3C\,120} had ceased.  In subsequent years the
outbursts have been weaker, and like other quasars and AGN individual
outbursts overlapped in both frequency and time.

\section{The Inverse Compton Catastrophe and Superluminal Motion}

As \Index{3C\,120} is relatively nearby, it did not present any particular
energetic problems.  Other sources, such as \Index{3C\,279} appeared to require
enormous energy, with up to $10^{60}$erg in the form of relativistic
particles generated in less than one year (Pauliny-Toth \& Kellermann
1966).  Moreover, causality arguments placed limits on the size and
corresponding radiation density of these transient sources.  The
amplitude and variability time scales of the observed outbursts
appeared to be inconsistent with the calculated energy losses due to
inverse Compton scattering (Hoyle, Burbidge, \& Sargent 1966).  It was
particularly difficult to understand the variations that we and others
observed at the longer wavelengths (Pauliny-Toth \& Kellermann
1966; Fanti 1979; Hunstead 1972).  These issues were exacerbated with
the discovery of even more rapid intraday variability (IDV) (see
Jauncey et al., these proceedings, page 199).  
Although now both the low frequency variablity
and IDV are thought to arise from scattering in the ISM, the required
dimensions still strain conventional theory.

The discovery of superluminal motion (Cohen et al.\ 1971, Whitney et
al. 1971) offered a simple solution in terms of relativistic beaming
(Cohen et al.\ 1977) following ideas suggested earlier by Rees (1966;
1967) and Woltjer (1966).  However, recently, more extensive
observations show little or no direct relation between variability
time scales and observed velocity for individual sources; but, there
remain ambiguities in the definition of the variability time scales.
It is encouraging, however, that the distribution of observed
velocities and variability time scales is consistent with a realistic
distributions of Lorentz factors (Cohen et al., 
these proceedings, page 177).

\section{$T_{\rm b}$(max) and Inverse Compton Cooling}

Kellermann \& Pauliny-Toth (1969) expressed the inverse Compton
cooling arguments in terms that could be directly related to
observations.  We showed that for any realistic magnetic field
strength that the ratio of inverse Compton radiation to synchrotron
radiation is simply a function of the peak brightness temperature,
$T_{\rm b}$.  We pointed out that if $T_{\rm b}>10^{12}$\,K, the relativistic
electron population would be quickly quenched by inverse Compton
cooling.  Early VLBI observations appeared to indicate that for the
compact flat spectrum radio sources, $T_{\rm b} \sim 10^{11-12}$\,K.
However, with the benefit of 30 years of hindsight, both the
derivation and the apparent agreement with VLBI observations were due
to a series of fortuitous accidents.

Ivan and I did not intend to discuss the brightness
temperature of synchrotron sources.  Rather, we wanted to show from
our multi-frequency flux density observations, that the flat spectra
observed for many sources actually consisted of multiple peaks and
valleys; and that this was due to the superposition of multiple
components each of which becomes optically thick at different
wavelengths.

This was in the very early days of VLBI, but multi-baseline multi
frequency observations were already showing that, as expected from
synchrotron theory, that components with the highest self absorption
cutoff frequency, had the smallest angular diameter (Kellermann et
al. 1969).  Partly as a test of synchrotron radiation models,
and partly to see if we could determine the magnetic field strength in
the individual components, we plotted the spectral cutoff frequency,
$\nu_{\rm m}$ vs. $S/\theta^{2}$.  As shown in Figure 1, we also plotted
lines of constant brightness temperature.  Using the relation between
angular size, surface brightness, and magnetic field, we showed
lines of constant magnetic field strength, and we noted
that the observed angular size and self absorption cutoff frequencies
were consistent with $B \sim 10^{-5}$ to $10^{-4}$\,Gauss.  We
were surprised to see that the observed values of $T_{\rm b}$ were
essentially all in the narrow range $10^{11}$ to $10^{12}$\,K.

\begin{figure}[htb!]
\begin{center}
\vspace{-10pt}
\includegraphics[clip,width=0.80\textwidth]{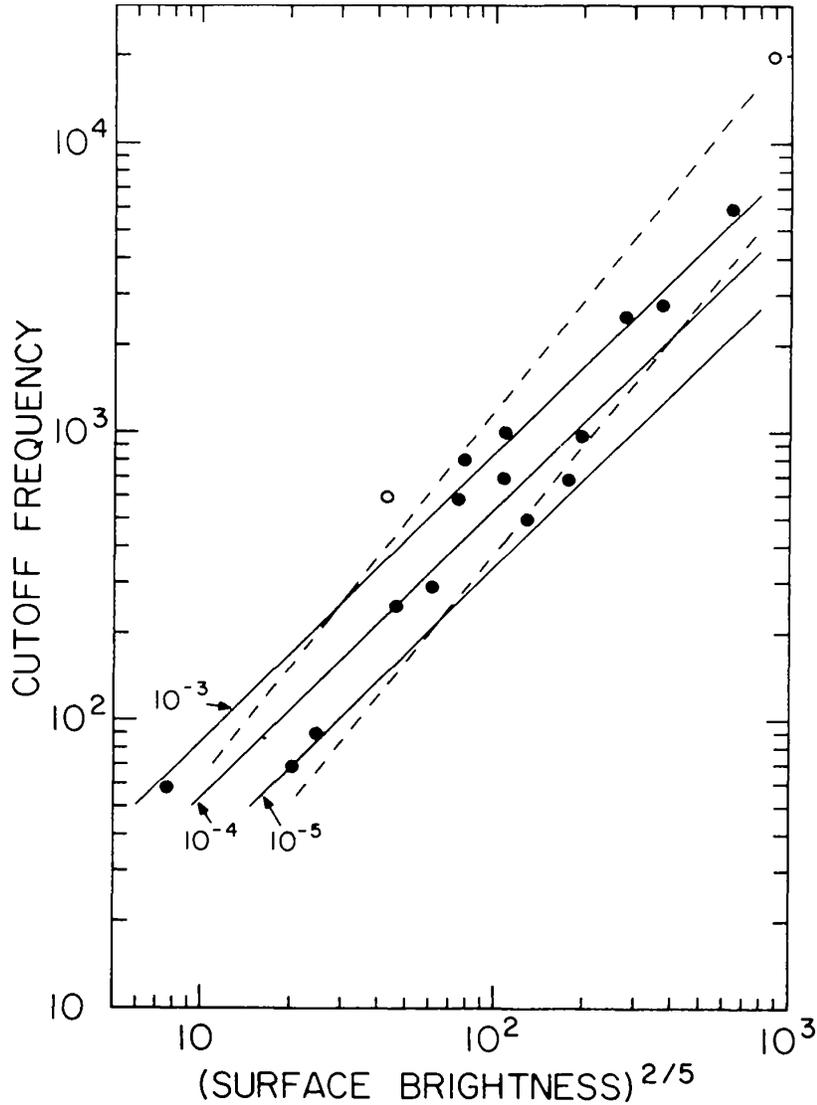}
\vspace{-15pt}
\end{center}
\caption{Spectral cutoff frequency (MHz) vs. $\mathrm{(surface \:
brightness)}^{2/5}$ for a number of opaque sources or components of
sources. The solid parallel lines show the expected dependence for
different values of the magnetic field.  The upper and lower dashed
lines represent constant brightness temperatures of $10^{11}$ and
$10^{12}$\,K respectively.  \textsl{Taken from Kellermann \& 
Pauliny-Toth 1969, ApJ, 155, L71.)}}
\end{figure}


We spent several weeks trying to understand what was special about
$T_{\rm b} \sim 3 \times 10^{11}$\,K.  Finally, we realized that with a
simple substitution of variables the ratio of inverse Compton
scattering to synchrotron radiation, $L_{\rm c}/L_{\rm s}$ can be expressed
simply in terms of the observed brightness temperature.  Hoyle,
Burbidge, \& Sargent (1966)
had noted that if $L_{\rm c}/L_{\rm s}>1$, then the effect of inverse
Compton cooling diverges, so we realized that values of
$T_{\rm b}>10^{12}$\,K cannot be sustained.

With the benefit of hindsight, the agreement between our derivation of
$T_{\rm b}$(max)=$10^{12}$\,K and the apparent maximum observed brightness
temperature was fortuitous and was the result of the following
over-simplifications and coincidences.

1) The limiting brightness temperature depends on the electron
energy distribution, the upper energy limit, and the detailed
geometry.  We made our calculations for a uniform slab geometry as
that was the only geometry for which we could handle the math, and
reported that $T_{\rm b}$(max) is in the range $10^{11-12}$\,K, not
$10^{12}$\,K as is often stated in the literature.  Based on more
detailed calculations, Readhead (1994) estimates $3 \times 10^{11}$\,K
as the limiting brightness temperature due to inverse Compton  cooling.

2)  There is no reason to assume that the VLBI observations were
made at the peak of the spectrum.  Below the peak, the flux density
falls off rapidly as $\nu^{2.5}$.
Above, the cutoff frequency, the brightness temperature also falls off
rapidly as $\nu^{-2.7}$.  Thus the observed brightness temperature is
likely to be less than the actual peak value.

3)  The early VLBI observations used in our study had only a limited
number of visibility measurements which we interpreted in terms of
simple circular Gaussian components.  We now know, of course, that the
true structure is more complex, usually with multiple separated
components.  Thus, the apparent sizes that we used to calculate $T_{\rm b}$
reflected the component separation rather than the size of any
individual component, and so it should not  be used to calculate $T_{\rm b}$.

4)  The inverse Compton limit to the maximum observed brightness
temperature refers only to a source at rest.  Doppler boosting
increases the observed brightness temperature by the Doppler factor,
$\delta$.  Most of the sources which we included in our 1969 analysis
are now known to show superluminal motion, (see Zensus et al., 
these proceedings, page 27).
These sources have significant Doppler factors, so
their brightness temperature can significantly exceed the inverse
Compton limit.

5)  There is growing evidence that that in some sources, at least,
the maximum brightness temperature is limited by free-free absorption
from an intervening ionized medium rather than by synchrotron self
absorption, (e.g., Vermeulen et al.\ 2003; Shaffer, Kellermann, \& 
Cornwell 1999;
Lister, these proceedings, page 71).
However, these sources also have measured
angular sizes which are consistent with self absorption, so that there
appears to be a mixture of SSA and FFA taking place.  It is curious
however, that in these sources, the FFA and SSA opacities appear to be
comparable.

6)  Coherent processes, which can increase the observed brightness
temperature above $10^{12}$\,K may be important (e.g., Melrose 2002).

7)  Synchrotron radiation from ultra relativistic protons can reach
a peak brightness temperature which is about a factor of
$(m_{\rm p}/m_{\rm e})^{9/7} \sim 10^{4}$ higher than from electrons
(Kardashev 2000).  However, in this case, the required magnetic field
is much larger by a factor of $(m_{\rm p}/m_{\rm e})^2 $ which is probably
unrealistic.

8)  Pauliny-Toth \& Kellermann (1966) noted that for a source at
rest, causality arguments indicated that the energy in relativistic
electrons greatly exceeds that in the magnetic field and would be
unrealistically large.  The effect of Doppler boosting is to increase
the apparent angular size and magnetic field strength, and so lower
the relativistic particle energy requirements by $ \delta^{-7/2}$.
Doppler factors $\delta \sim 10$ will typically bring the particle and
magnetic energies into equilibrium.  In this case the the peak
brightness temperature is reduced by about a order of magnitude over
the inverse Compton limit (Readhead 1994).

9) As we pointed out in our 1969 paper (see also Slysh (1992), the
inverse Compton limit only applies after the source has reached
equilibrium.  For a limited period following the injection of a new
supply of relativistic electrons, or if there is continuous
reacceleration, brightness temperatures may be observed which are in
excess of the inverse Compton limit.  In this case, copious X-ray
emission should be observed.  For example first order Compton
scattering will allow $T_{\rm b}$ to exceed $10^{12}$\,K for about one year
following the injection of relativistic electrons.  Higher order
scattering will reduce the lifetime further, but the scattering is
ultimately limited by quantum effects.

10)  Finally, we point out that the measured values of brightness
temperature which prompted our discussion of $T_{\rm b}$(max) and inverse
Compton cooling was merely the natural consequence of the size of the
Earth and the limited range of flux density observed, and had nothing
to do with radio source physics. This is because the resolution of any
interferometer is given by $\theta = \lambda /D$ where $\lambda$ is
the wavelength of observation, and $D$ is the baseline length.  For a
radio source of flux density $S$ and angular size, $\theta$, the
brightness temperature $T_{\rm b} = 2k\lambda^{2}S/\theta^{2}$.  Thus
$T_{\rm b} \propto SD^{2}$, and is independent of angular size and
observing wavelength.  For $1<S_{\rm Jy}<10$, and $D \sim 5000$\,km, $T_{\rm b}
\sim 10^{11-12}$\,K.

\section{Current Issues}

More than three decades have passed since Ivan and I considered the
question of a limiting brightness temperature.  In many ways we are
still asking the same questions about the same issues.  However, due
to tremendous advances in both the observational data and theoretical
models, our questions are at a much higher level than before.
Although the
observations sometimes strain conventional synchrotron models and
relativistic boosting, there does not appear to be any show
stoppers.  However, the number of free parameters needed to accommodate the
observations has become uncomfortably large, and it has proven
difficult to confirm by observation the predicted effects of
relativistic motion.  The early models make simple predictions.  But,
today, consideration of jet instead of ballistic flows, possible
differences between the bulk velocity and the pattern velocity, combined with
a distribution of intrinsic Lorentz factors has removed the
predictability, and thus the simplicity, if not the attractiveness of
these models.

It is perhaps sobering to recall that the concept of relativistic
beaming was first introduced in
the mid 1960's to explain the rapid variability that was
observed in quasars, and which appeared to be dramatically confirmed
by the discovery of superluminal motion in 1971.  But, predictions of
the distribution of observed velocity distribution can only be
reconciled with observations by introducing modifications to the
simple theory which involves new free parameters. Meanwhile simple 
non-Doppler boosting models such as suggested by Ekers and Liang (1989)
are surprisingly consistent with the observed velocity distribution.
The main evidence for Doppler boosting remains, as in the 1960's, the
rapid flux density variations, not superluminal motion.

As discussed by Cohen et al.\ 
(these proceedings, page 177)
application of the
standard causality argument combined with the measured velocities can
be used to estimate the intrinsic brightness temperature and Doppler
factor.  Although there is no simple direct relationship between the
brightness temperatures calculated from variability time scales and
measured velocities, the observed values are consistent with a
reasonable distribution of Lorentz factors.  However ambiguities in
defining the variability time scales, in relating the variable
component to the moving component, and the apparent differences
between the bulk velocity and pattern velocity may preclude any
meaningful quantitative conclusions.

Modern VLBA (Kellermann et al; 1998, Zensus et al.\ 2002; Kovalev,
these proceedings, page 65)
and VSOP space VLBI (Tingay et al.\ 2001)
observations suggest maximum observed brightness temperatures in the
range $10^{11-13}$, much as we measured in the first years of
VLBI. Although VSOP uses longer baselines, fringe visibilities are more
accurately determined on the VLBA so the corresponding size limits for
unresolved components are comparable for the two instruments.  The
Russian \textsl{RadioAstron} space VLBI mission which is scheduled for
launch in 2006 will give more than an order of magnitude increase in
baseline length.  This will directly measure brightness temperatures
up to $10^{15}$\,K and perhaps allow better comparison with calculated
Doppler factors.

The maximum relativistic boosting occurs when the the motion
is oriented directly along the line of sight.  In this case,
$T_{\rm b}$(obs) $ \sim 2\gamma T_{\rm b} $(int)   where $\gamma$ is the Lorentz
factor.  The observed range of $T_{\rm b}$ is consistent with the observed
range of apparent velocity and intrinsic values of $T_{\rm b}\sim
10^{10-11}$\,K, or close to both the inverse Compton limit and
the equilibrium value.  However, perhaps surprisingly, there is no
simple one-to-one relation between the observed brightness temperature
and measured superluminal motions.

Using the same kind of causality arguments used in the 1960s to
``predict'' superluminal motion, IDV observations suggest brightness
temperatures as much as $\sim 10^{21}$\,K  they are intrinsic and
$T_{\rm b} \sim 10^{15}$\,K if the variations are due to interstellar
scintillations as seems likely, (Jauncey et al., 
these proceedings, page 199).
It may
be hard to reconcile these numbers with reasonable Doppler factors.

In principle, the intrinsic brightness temperature and Doppler factor
can be determined from the ratio of observed inverse Compton X-ray
flux density to the radio flux density from synchrotron radiation,
(e.g., Unwin et al.\ 1997).  However, there are three problems in
implementing this technique.  1) It is difficult to get
contemporaneous radio and X-ray light curves.  2) We don't know
whether the X-rays are coming from the core or from somewhere along
the jet. 3) There may be an unknown additional (non-Compton), amount
of X-ray emission for example from the hot ionized torus surrounding
the central black hole.

The concept of equipartition in extragalactic radio sources was first
introduced by Burbidge (1959) to minimize the total energy content of
extended lobe radio sources which typically have ages of $10^{8-9}$
years.  We do not know if it is appropriate to assume that
equipartition conditions apply in the short lived compact cores
and jets.  Indeed, the presence of bulk relativistic motion in the
parsec scale jets suggests that the plasma may not be contained by the
magnetic field so that there may be an excess of particle energy over
magnetic energy.  Comparison of Doppler factors deduced from
variability time scales and measured proper motions suggests that the
intrinsic brightness temperatures may be closer to the equipartition
value than to the inverse Compton limit (Cohen et al., 
these proceedings, page 177),
but the difference between the two is small and perhaps
not meaningful.

\begin{acknowledgements}
I am grateful to the many colleagues with whom I have had the
privilege of working; especially to Ivan Pauliny-Toth who could
not be here with us, but who collaborated with me for many years in
the work described in this paper; as well as Marshall Cohen, who has
been a valued colleague and good friend for nearly all of the 40~years
that I have been working in radio astronomy, and who organized this
meeting.  Special thanks go to Dan Homan and Matt Lister who have
tried to educate me in the subtleties of the \LaTeX\ language.  The NRAO
is is a facility of the NSF which is operated under cooperative
agreement by AUI.
\end{acknowledgements}

\printindex

\end{document}